\documentclass{sigchi}

\usepackage{balance}  % to better equalize the last page
\usepackage{graphics} % for EPS, load graphicx instead
\usepackage{times}    % comment if you want LaTeX's default font
\usepackage{url}      % llt: nicely formatted URLs
\usepackage{graphicx}
\usepackage{pbox}
\usepackage{caption}
\usepackage{subcaption}

\makeatletter
\def\url@leostyle{%
  \@ifundefined{selectfont}{\def\UrlFont{\sf}}{\def\UrlFont{\small\bf\ttfamily}}}
\makeatother
\def\pprw{8.5in}
\def\pprh{11in}

\usepackage[pdftex]{hyperref}
\hypersetup{
pdftitle={U.S. Religious Landscape on Twitter},
pdfauthor={Anonymous},
pdfkeywords={religion; Twitter},
bookmarksnumbered,
pdfstartview={FitH},
colorlinks,
citecolor=black,
filecolor=black,
linkcolor=black,
urlcolor=black,
breaklinks=true,
}

\newcommand\T{\rule{0pt}{1.2ex}}       % Top strut
\newcommand\B{\rule[-1.2ex]{0pt}{0pt}} % Bottom strut

\begin{document}

\title{U.S. \ Religious Landscape on Twitter}

\numberofauthors{3}
\author{
  \alignauthor Lu Chen\thanks{This work was done while the first author was an intern at Qatar Computing Research Institute.}\\
%  \thanks{This work was done while the first author was an intern at Qatar Computing Research Institute.}
    \affaddr{Kno.e.sis Center, \\ Wright State University}\\
    \affaddr{Dayton, OH, USA}\\
    \email{chen@knoesis.org}\\
  \alignauthor Ingmar Weber\\
    \affaddr{Social Computing, Qatar Computing Research Institute}\\
    \affaddr{Doha, Qatar}\\
    \email{iweber@qf.org.qa}\\   
 \alignauthor Adam Okulicz-Kozaryn\\
    \affaddr{Rutgers-Camden}\\
    \affaddr{Camden, NJ, USA}\\
    \email{adam.okulicz.kozaryn@gmail.com}\\
}

\maketitle

\begin{abstract}
Religiosity is a powerful force shaping human societies, affecting domains as diverse as economic growth or 
the ability to cope with illness. As more religious leaders and organizations as well as believers 
start using social networking sites (e.g., Twitter, Facebook), online activities become important extensions to traditional religious rituals and practices. 
However, there has been lack of research on religiosity in online social networks. This paper takes a step toward the understanding of several important aspects of religiosity on Twitter, based on the analysis of more than 250k U.S.\ users who self-declared their religions/belief, including \emph{Atheism}, \emph{Buddhism}, \emph{Christianity}, \emph{Hinduism}, \emph{Islam}, and \emph{Judaism}. 
Specifically, (i) we examine the correlation of geographic distribution of religious people between Twitter and offline surveys. 
(ii) We analyze users' tweets and networks to identify discriminative features of each religious group, and explore supervised methods to identify believers of different religions. (iii) We study the linkage preference of different religious groups, and observe a strong preference of Twitter users connecting to others sharing the same religion.
\end{abstract}

\section{Introduction}
Religiosity is a powerful force shaping human societies, and it is persistent -- 94\% of Americans believe in God and this percentage has stayed steady over decades \cite{sedikides10}. It is important to study and understand religion because it affects multiple domains, ranging from economic growth \cite{barro03al}, organizational functioning \cite{day2005religion} to the ability to better cope with illness \cite{campbell76etal}. 
A key feature of any belief system such as religion is replication -- in order to survive and grow, religions must replicate themselves both vertically (to new generations) and horizontally (to new adherents). The Internet already facilitates such replication. Traditional religions are likely to adapt to the societal and historic circumstances and take advantage of social media. Many churches and religious leaders are already using social networking sites (e.g., Twitter, Facebook) to connect with their believers. While social networking and social media become important means of religious practices, our understanding of religiosity in social media and networking sites remains very limited. In this paper, we take a step to  bridge this gap by studying the phenomenon of religion for more than 250k U.S.\ Twitter users, including their tweets and network information.

Twitter, because of its global reach and the relative ease of collecting data, is becoming a great treasure trove of information for computer and social scientists. Researchers have studied various problems using Twitter data, such as mood rhythms \cite{golder09}, happiness\cite{dodds10}, electoral prediction \cite{chen2012twitter}, or food poisoning \cite{cox14}. However, studies that explore the phenomenon of religion in social networking sites are still rare so far. To date, the most relevant study investigates the relationship between religion and happiness on Twitter \cite{ritteretal14spps}. It examines the difference between Christians and Atheists concerning the use of positive and negative emotion words in their tweets, whereas our work focuses on the religiosity of Twitter users across five major religions and Atheism. One recent study \cite{Nguyen2014religion} addresses the prediction of users' religious affiliation (i.e., Christian or Muslim) using their microblogging data, which focuses on building the classification model but not studying the phenomenon.

We collected U.S. Twitter users who self-reported their religions as \emph{Atheism}, \emph{Buddhism}, \emph{Christianity}, \emph{Hinduism}, \emph{Islam}, or \emph{Judaism} in their bios, and further collected their tweets and friends/followers. Our dataset comprises 250,840 U.S. Twitter users, the full lists of their friends/followers, and 96,902,499 tweets. In particular, we explore the following research questions in this paper: 

\begin{enumerate}
\item \emph{How does the religion statistics on Twitter correlate with that in the offline surveys?} Our correlation analysis shows that: (1) There is a moderate correlation between survey results and Twitter data regarding the distribution of religious believers of a given denomination across U.S.\ states, e.g., the macro-average Spearman's rank correlation of all the denominations is $\rho = .65$. (2) Similarly, the fraction of religious people of any belief within a given U.S.\ state in surveys matches well with that of Twitter users referencing any religion in their profiles with a Pearson Correlation of $r = .79$ ($p < .0001$). 
\item \emph{Whether or not do various religious groups differ in terms of their content and network? Can we build a classifier to accurately identify believers of different religions?} Specifically, (1) By looking at discriminative features for each religion, we show that users of a particular religion differ in what they discuss or whom they follow compared to random baseline users. (2) We build two classifiers that detect religious users from a random set of users based on either their tweets' content or the users they follow, and we find that the network ``following'' features are more robust than tweet content features, independent of the religion considered. 
\item \emph{Does the in-group linkage preference exist in any particular religious denomination?} Our main findings include: (1) We find strong evidence of same-religion linkage preference that users of a particular denomination would have an increased likelihood to follow, be-followed-by, mention or retweet other users of the same religion. For example, our results show that following someone of the same religion is $646$ times as likely as following someone of a different religion based on a macro-average of six denominations. (2) We show that ``the pope is not a scaled up bishop'' in that hugely popular religious figures on Twitter have a
higher-than-expected share of followers without religious references in their profiles.
\end{enumerate}

Our findings may not only improve the understanding of religiosity in social media but, as the Internet is becoming a medium for religious replication, also have implications for religion \emph{per se}.

\section{Related Work}
Religion shapes human society and history, and defines a person in many ways. It has such a large influence on people that it can be used as a measure of culture \cite{iannaccone98}. Fundamentally, religiosity satisfies ``the need to belong'', that is,
people who are religious and live in religious societies, feel that they are part of that society \cite{aokrel,okulicz11}. 
Religiosity does predict multiple outcomes such as economic growth, happiness, trust, and cooperation \cite{barro03al,campbell76etal,aokrel,okulicz11,sosis05,hall10}. 

In the past few years, religion has been the subject of some Social Informatics research, particularly examining the role of Internet-based technologies in religious practice \cite{wyche2006technology,wyche2008sun,ho2008muslim}. For example, Wyche et al. \cite{wyche2006technology} explore how American Christian ministers have adopted technologies such as the World Wide Web and email to support the spiritual formation and communicate with their laity. In another study \cite{wyche2008sun}, researchers discuss the design and evaluation of a mobile phone application that prompts Muslims to their five daily prayer times. There is a study of ``church'' (and ``beer'') mentions on Twitter, which corroborates our results showing more religiosity in South Eastern states\footnote{http://www.floatingsheep.org/2012/07/church-or-beer-americans-on-twitter.html}. 
In addition, by examining how religious people use various technologies (e.g., home automation technology, information and communications technology) for their religious practices, and whether that is different from their secular counterparts, implications can be gained to guide the future design of technologies for religious users \cite{woodruff2007sabbath,wyche2008re,wyche2009extraordinary}. Another line of research in this context investigates the process of ``spiritualising of Internet'' -- how religious users and organizations shape and frame the Web space to meet their specific needs of religious rituals and practices \cite{campbell2005spiritualising,busch2011come}. It is also suggested by some researchers that studying religion on the Internet provides a microcosm for understanding Internet trends and implications \cite{campbell2013religion}.

Some other studies have focused on online religious communities. For example, McKenna and West \cite{mckenna2007give} conduct a survey study of the online religious forums where believers interact with others who share the common faith \cite{mckenna2007give}. Lieberman and Winzelberg \cite{lieberman2009relationship} examine religious expressions within online support groups on women with breast cancer. It is reported that the same self and social benefits (e.g., social support, emotional well-being) found to be associated with the involvement in traditional religious organizations can also be gained by participation in online religious communities. 

% religion in social network
While much research effort has been made to understand religious use of Internet technologies, we know very little about religiosity in online social networks. On the other hand, there is recently an explosion of studies on Twitter \cite{golder09,dodds10,miller11,cox14,kwak2010twitter,de2010birds,kang2012using}, yet we do not know much specifically about religiosity on Twitter. Wagner et al.\ \cite{wagner2013religious} develop classifiers to detect Twitter users from different categories, including category \textit{religious}; Nguyen and Lim \cite{Nguyen2014religion} build classifiers to identify Christian and Muslim users using their Twitter data, but neither of the two studies addresses the analysis of the phenomenon of religion on Twitter. \cite{ritteretal14spps} appears to be the most relevant study, which focuses on exploring the relationship between religion and happiness via examining the different use of words (e.g., sentiment words, words related to thinking styles) in tweets between Christians and Atheists. Our present work differs both in scope and purpose.

\section{Data}
\label{data}
Identifying religiosity on Twitter is non-trivial as users can belong to a particular religious group without making this affiliation public on Twitter. In this section we describe how we collect data, with a general focus on precision rather than recall, and how we validate the collected data.
Concerning the selection of religions we decided to limit our analysis to the world's main religions, concretely, \emph{Buddhism}, \emph{Christianity}, \emph{Hinduism}, \emph{Islam}, and \emph{Judaism}. We also included data for \emph{Atheism}, and an ``\emph{undeclared}'' baseline set of users. We focused our data collection on the U.S.\ as this allowed us to obtain various statistics about the ``ground truth'' distribution of religions across U.S.\ states.

The advantage of Twitter is that data are captured unobtrusively (free from potential bias of survey or experimental setting). However, Twitter has its own biases and the issues of representativeness need to be taken into account when interpreting the results. For example, according to a study\footnote{http://www.beevolve.com/twitter-statistics/} published in 2012, Twitter users are predominantly young (74\% fall between 15 to 25 years of age). It is reported in another study \cite{mislove11} in 2011 that Twitter users are more likely to be males living in more populous counties, and hence sparsely populated areas are underrepresented; and race/ethnicity is biased depending on the region.

\subsection{Data Collection and Geolocation}
\label{dataCollection}

\begin{table*}[t]
\centering
\scalebox{0.8}{
\begin{tabular}{|l|l|l|l|l|l|l|l|}
\hline
User group                            & Atheist & Buddhist & Christian & Hindu  & Jew & Muslim & Undeclared \\ \hline
\# of users                           & 7,765    & 2,847     & 202,563    & 204    & 6,077   & 6,040   & 25,344      \\ \hline
Mean \# of tweets per user             & 3976.8  & 2595.7   & 1981      & 2271.5 & 2095.7 & 3826.5 & 1837.3     \\ \hline
Mean \# of tweets per user per day     & 3.3     & 2        & 1.8       & 1.9    & 1.8    & 4.2    & 1.9        \\ \hline
Stdev of \# of tweets per user per day & 8.3     & 5.8      & 5.3       & 4.8    & 6      & 9.2    & 5.8        \\ \hline
Median of \# of friends               & 179     & 144      & 151       & 119    & 163    & 166    & 114        \\ \hline
Mean \# of friends per user           & 442.8   & 452.3    & 370       & 277.2  & 399.6  & 344.1  & 295.5      \\ \hline
Stdev of \# of friends per user       & 659     & 1825.4   & 2179.6    & 470.1  & 882.9  & 243.1  & 991.6      \\ \hline
Median of \# of followers             & 79      & 77       & 77        & 74     & 112    & 104    & 52         \\ \hline
Mean \# of followers per user         & 707.5   & 628.9    & 418.2     & 308.2  & 665.2  & 467.9  & 400        \\ \hline
Stdev of \# of followers per user     & 23987.4 & 4873.6   & 6834.6    & 889.8  & 5063.2 & 2855   & 6691.6     \\ \hline
\end{tabular}
}
\caption{{\footnotesize Description of the dataset.}}
\label{dataOverview}
\end{table*}

To obtain a list of users who are most likely believers of the six denominations of interest, we search Twitter user bios via
Followerwonk\footnote{https://followerwonk.com/bio} with a list of keywords\footnote{We realize that this keyword list is not complete (e.g.\ Mormons self-identify as Christians) of these denominations, and leave it for the future research to explore an extended list. Our current focus is on precision, with a potential loss in recall.}. From Followerwonk, we obtain these users' screen names, with which we collect more information of these users through Twitter API, including their self-declared locations, descriptions, follower counts, etc. 

In addition, we collect another group of Twitter users who do not report any of the above mentioned religions/beliefs in their bios. Specifically, we generate random numbers as Twitter user IDs \footnote{We registered a new Twitter account and obtained its ID, then we generated random numbers ranging from 1 to that ID, i.e., 2329304719. Note that Twitter IDs are assigned in ascending order of the time of account creation.}, collect these users' profiles via Twitter API, and remove the users who appear in any of the user collections of the six denominations from this set. We label this user group as \emph{Undeclared}.

We then identify users from the United States using users' self-declared locations. We build an algorithm to map location strings to U.S.\ cities and states. The algorithm considers only the locations that mention the country as the U.S.\ or do not mention any country at all, and uses a set of rules to reduce incorrect mappings. For example, ``IN'' may refer to the U.S.\ state ``Indiana'' or be a part of a location phrase, e.g, ``IN YOUR HEART''. To avoid mapping the latter one to ``Indiana'', the algorithm considers only the ones where the token ``IN'' is in uppercase, and mention either the country U.S.\ or a city name. If a city name is mentioned without specifying a state, and there are more than one states that have a city named that, the algorithm maps it to the city and state which has the largest population. 

We keep only the users whose location string is mapped to one of the 51 U.S. states (including the federal district Washington, D.C.), the language is specified as ``en'', the self-description bio is not empty\footnote{This only happened for the undeclared users as the other users were found by searching in their bio. We removed such users with a empty bio as they were likely to have a very different activity pattern than users providing information about themselves.}, and tweet count is greater than $10$.  Overall, this dataset contains 250,840 users from seven user groups. Using Twitter API, we also obtain the collection of tweets (up to 3,200 of a user's most recent tweets as by the API restrictions), and the list of friends and followers of these users. Table \ref{dataOverview} provides an overview of the dataset. If we measure the active level of users in terms of the number of tweets, friends and followers, on average, \emph{Atheists} appear to be more active than religious users, while the \emph{Undeclared} group generally appears to be less active than other groups. Among the five religious groups, \emph{Muslim} users have more tweets, both \emph{Muslim} and \emph{Jew} users tend to have more friends and followers, compared with other religions.      

It is important to note that only the Twitter users who publicly declare their religion/belief in their bios are included in our data collection, while vast majority of believers may not disclose their religion in their Twitter bios and thus not included. This may lead to bias toward users who are very religious or inclined to share such information.

\subsection{Data Validation}
Mentioning a religion-specific keyword (e.g., ``Jesus'') in the bio may not necessarily indicate the user's religious belief. Table~\ref{bioExample} shows example user bios including both true positives (religion/belief is correctly identified) and false positives (religion/belief is not correctly identified). To evaluate the quality of our data collection, we randomly selected 200 users from each user group, and manually checked their bios and religion labels. The precision of religion identification is represented as $\frac{\# true \ positive}{\# total}$. Overall, macro-averaged precision across all the groups is $0.91$, which shows that our way of identifying religiosity is quite precise. The identification of Jewish users is found to be the least accurate ($0.78$), because it contains the largest fraction of false positives (mostly indicating opposition and hatred) as illustrated by Examples~7 and 8 in Table~\ref{bioExample}\footnote{We chose not to show offensive profile examples here. Disturbing examples can, however, be easily found using \url{http://followerwonk.com/bio/}.}. Sadly, ``digital hate'' seems to be on the rise \cite{wiesenthal}.

\begin{table}[ht]
\centering
\scalebox{0.9}{
\begin{tabular}{|l|l|}
\hline
1  & Animal lover.Foodie.Model.\textbf{Buddhist}.                                                                                                                  \\ \hline
2  & \pbox{20cm}{\textbf{Atheist}, Doctor Who fan, the left side of politics, \\ annoyed by happy-horseshit \& pseudo-spiritual people \T\B}                                            
\\ \hline
3  & \textbf{ISLAM} 100\% \T\B                                                                                                                                    \\ \hline
4  & \pbox{20cm}{a little bit cute,a loving sis,a good follower of \textbf{jesus},.,..\\ a friendly one.. \T\B}                                                                       
\\ \hline
5  & \pbox{20cm}{\textbf{Christian}, Wife of @coach\_shawn10, Mother of 3 \\ beautiful daughters, Sports Fan, AKA. I'm blessed \\ and highly favored! \T\B}                             
\\ \hline
6  & \pbox{20cm}{Worked with The \textbf{Hindu} Business Line \& Dow Jones News-\\wires. Tracking/Trading Stock market for over 15 years. \T\B}                                       
\\ \hline
7  & \pbox{20cm}{PhD in Worthless Information. Surprisingly not \\ \textbf{Jewish} or Amish. We Are! Let's Go Buffalo! \T\B}                                                         
\\ \hline
8  & my boss is a \textbf{Jewish} Carpenter \T\B                                                                                                                        \\ \hline
9 & \pbox{20cm}{\textbf{JESUS}! I get paid to go to football games. Social \\ life? What is that? Follow @username for all things Sports. \\I think I'm funny, I'm probably wrong. \T\B}
\\ \hline
\end{tabular}
}
\caption{{\footnotesize Example user bios. Example 1-5 are true positive, and 6-9 are false positive.}}
\label{bioExample}
\end{table}

We also evaluate the geolocation results of the same data sample. The authors manually identified U.S.\ states from location strings of users in the sample. Among all the 1,400 users, 329 users' locations were mapped to U.S. states by the authors. The algorithm identified 298 U.S. users and mapped their locations to states, among which 289 were consistent with the manual mapping. The algorithm achieved a precision of $\frac{289}{298} = 0.97$ and a recall of $\frac{289}{329} = 0.88$.      

\section{Correlation Analysis of Religion Statistics between Twitter Data and Surveys}

In this section, we explore how religion statistics we observed in our Twitter dataset correlate with that in offline surveys. 

\subsubsection[Pew Research U.S.\ Religious Landscape Survey]{Pew Research U.S.\ Religious Landscape Survey\footnote{http://religions.pewforum.org/}}

By counting the Twitter users of each denominations for each state, we get estimates of the religious composition in each of the 51 states. Pew Research Religious Landscape Survey also provides the religious composition by U.S.\ states, which covers nine categories including Buddhist, Christian, Hindu, Jew, Muslim, Unaffiliated, Other World Religion, Other Faiths, and Don't know/refused. The Unaffiliated category includes Atheist, Agnostic, and Nothing-in-particular. Since our data collection does not include categories such as Other World Religion, Other Faiths, or Don't know/refused, and our group of Atheist does not include Nothing-in-particular category, we remove these categories that are not included in our data collection and recalculate the composition among the remaining ones. 

The per-value correlation across all the religions and states is $r > .995$, but since Christians are dominant in every state, it's easy to get a high correlation by just guessing Christian $= 100\%$ in every state. So we also conduct the correlation analysis of each religion across the 51 states. The Pearson's $r$ on Christian and Jew are $r = .73$ and $r = .77$ ($p < .0001$ in both cases), respectively. The Spearman's rank correlation on Christian, Jew and Buddhist are $\rho = .77$, $\rho = .79$ and $\rho = .75$ ($p < .0001$ in all three cases). But the correlations on Muslim and Hindu are only at $.15 < r < .30$ ($.03 < p < .3$) and $.48 < \rho < .50$ ($p < .0004$).

The proportions by denomination in our Twitter sample from Table~\ref{dataOverview} can also be compared with the actual proportions -- for instance according to Pew\footnote{http://religions.pewforum.org/reports} there are about twice as many Jews as Buddhists in the U.S., and our sample shows the same proportions; there are about 2 times more Buddhists than Hindus; yet our sample has 10 times more Buddhists than Hindus.

The most plausible reason for non-perfect fits, especially for the geographic distribution of Muslims and Hindus in the U.S., is simply that the Twitter population is a biased selection of the general population as explained in Section~\ref{data}. The sample size is another potential reason. Especially for small U.S.\ states we have only few non-Christian users in our set. Finally there are most likely also religion-specific differences in terms of the inclination to publicly state one's religious affiliation in a Twitter profile. 

\subsubsection[Gallup U.S. Religiousness Survey]{Gallup U.S. Religiousness Survey\footnote{http://www.gallup.com/poll/125066/State-States.aspx}} 

Gallup's survey measures religiousness based on respondents' self-reported importance of religion in their daily lives and their attendance at religious services \cite{gallupReligiousness}. The survey provides the proportions of \textit{very religious}, \textit{moderately religious} and \textit{non-religious} residents in each U.S.\ state. We get the percentage of religious residents by adding the very religious and moderately religious proportions together.

We count the number of religious Twitter users (including Buddhist, Christian, Hindu, Jew, and Muslim) in each state, which is $N^R(s)$, where $s$ can be any of the 51 U.S.\ states, e.g., $N^R(Ohio)$. By adding them together we get the total number of religious users in the U.S., i.e., $N^R(all)$. Then the fraction of religious users of state $s$ is $\frac{N^R(s)}{N^R(all)}$. In a similar way, we can get the fraction of undeclared users of state $s$ as $\frac{N^U(s)}{N^U(all)}$, where $N^U(s)$ is the number of undeclared users in state $s$, and $N^U(all)$ is the total number of undeclared U.S.\ users. Note that we do not differentiate users on Twitter according to degrees of religiosity for this study as, we believe, users that explicitly state their religious affiliation online are likely to be comparatively more religious.

\begin{figure*}[t!]
  \centering
  \begin{subfigure}[t]{0.48\textwidth}
        \includegraphics[width=\textwidth]{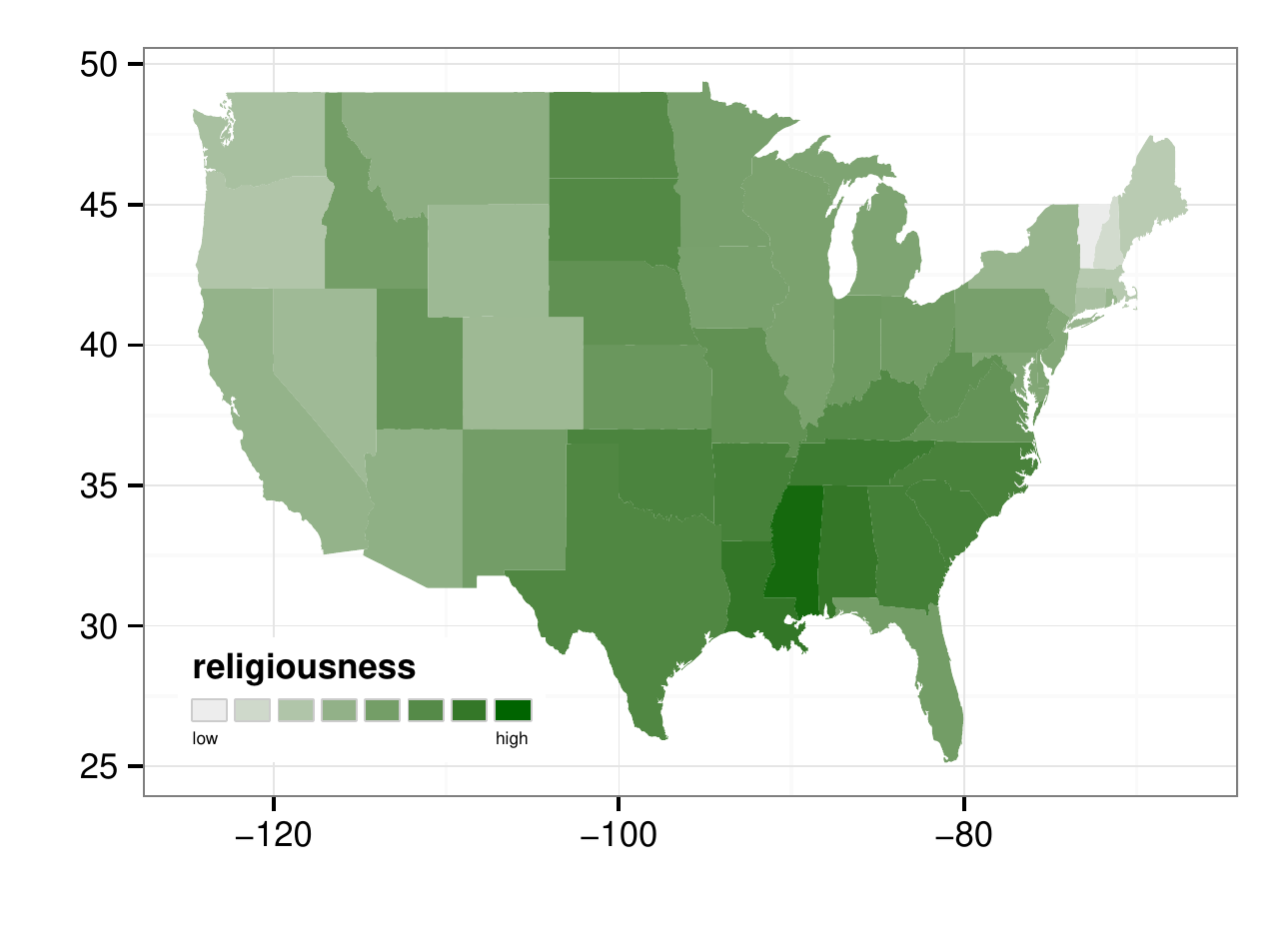}
        \caption{Gallup}
        \label{gallupRel}
  \end{subfigure}
  \begin{subfigure}[t]{0.48\textwidth}
        \includegraphics[width=\textwidth]{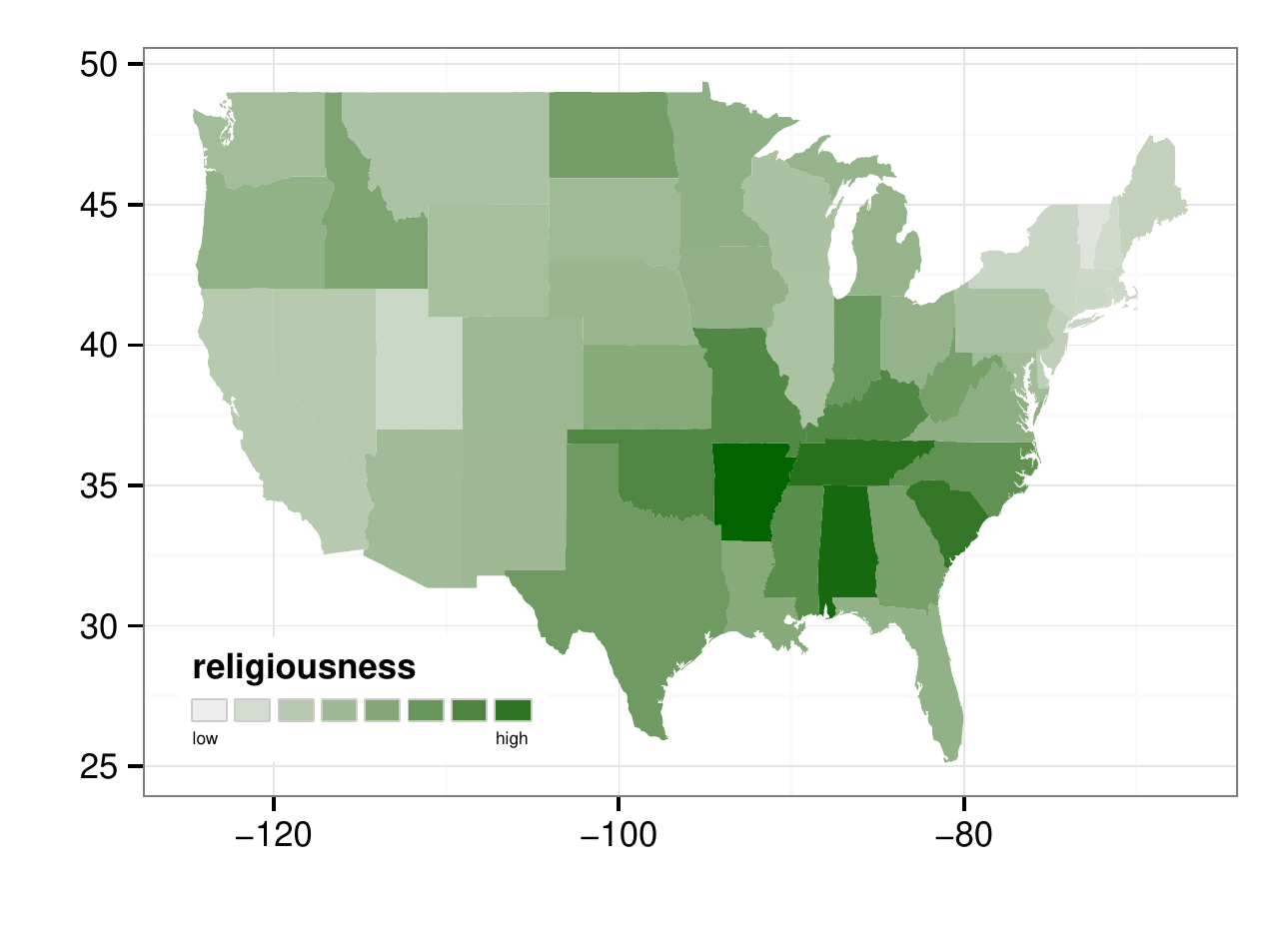}
        \caption{Twitter}
        \label{twitterRel}
  \end{subfigure}
 \caption{Map of State Variations of Religiousness in the U.S.}
  \label{religiousnessMap}
  \vspace{-1.0em}
\end{figure*}

Then we measure the religiousness of state $s$ as $\frac{N^R(s)}{N^R(all)} / \frac{N^U(s)}{N^U(all)}$. The higher the score, the more religious the state as it has a larger-than-expected number of Twitter users with a self-stated religious affiliation. Correlating this religiousness score per state against the Gallup survey shows a respectable fit of Pearson's $r = .79$ ($p < .0001$). Figure~\ref{religiousnessMap} shows state variations of religiousness by both the survey data and Twitter data. They agree on 11 of the top 15 most religious states (e.g., Alabama, Mississippi, and South Carolina) and 11 of the top 15 least religious states (e.g, Vermont, New Hampshire, and Massachusetts). However, Utah is the second most religious state according to Gallup survey, but is  one of the least religious states according to our data collection. The main reason might be that Mormonism (the dominant religion in Utah) is underrepresented in our dataset as we did not scan for related terms in the users' profiles.

In addition, the Pearson's $r$ between the number of undeclared users per state and the population of those states is $.986$ ($p < .0001$), which suggests a good level of representativeness in terms of the number of Twitter users. 

\section{Identification of Believers of Various Religious Denominations}
In this section, we explore the discriminative features of a religion that differentiate its believers from others, and build classifiers to identify religious Twitter users of various denominations. 

By exploring the features that are effective for identifying Twitter users of a certain religious denomination, we would gain insight on the important aspects of a religion. For example, the comparison of tweet content based features and network based features in a classifier would show whether it is more about ``the company you keep'' or ``what you say online'' that tells you apart from others of a different religious belief. In addition, by looking at how easy/difficult it is for a classifier to recognize believers of a particular religion, we could see which religions are ``most religious'' in that they differ most from ``normal'' behavior on Twitter. This is not just a classification question but also a societal question: religions that could be told easily by with whom you mingle (network) are probably more segregated, and possibly intolerant towards other groups -- in general religiosity and prejudice correlates \cite{hall10}. Again, this has broader societal implications because these linkage or group preferences are likely to be present in the real world as well -- for instance, real world traits and behaviors such as tolerance, prejudice, and openness to experience are likely to be correlated with our findings. For example, differences in hashtag usage between Islamists and Seculars in Egypt has been found to indicate ``polarization'' in society \cite{weberetal13asonam}.

\subsection{Discriminative Features}

\subsubsection{What do they tweet?}
\begin{figure*}[t!]
\centering
\includegraphics[width=0.8\textwidth]{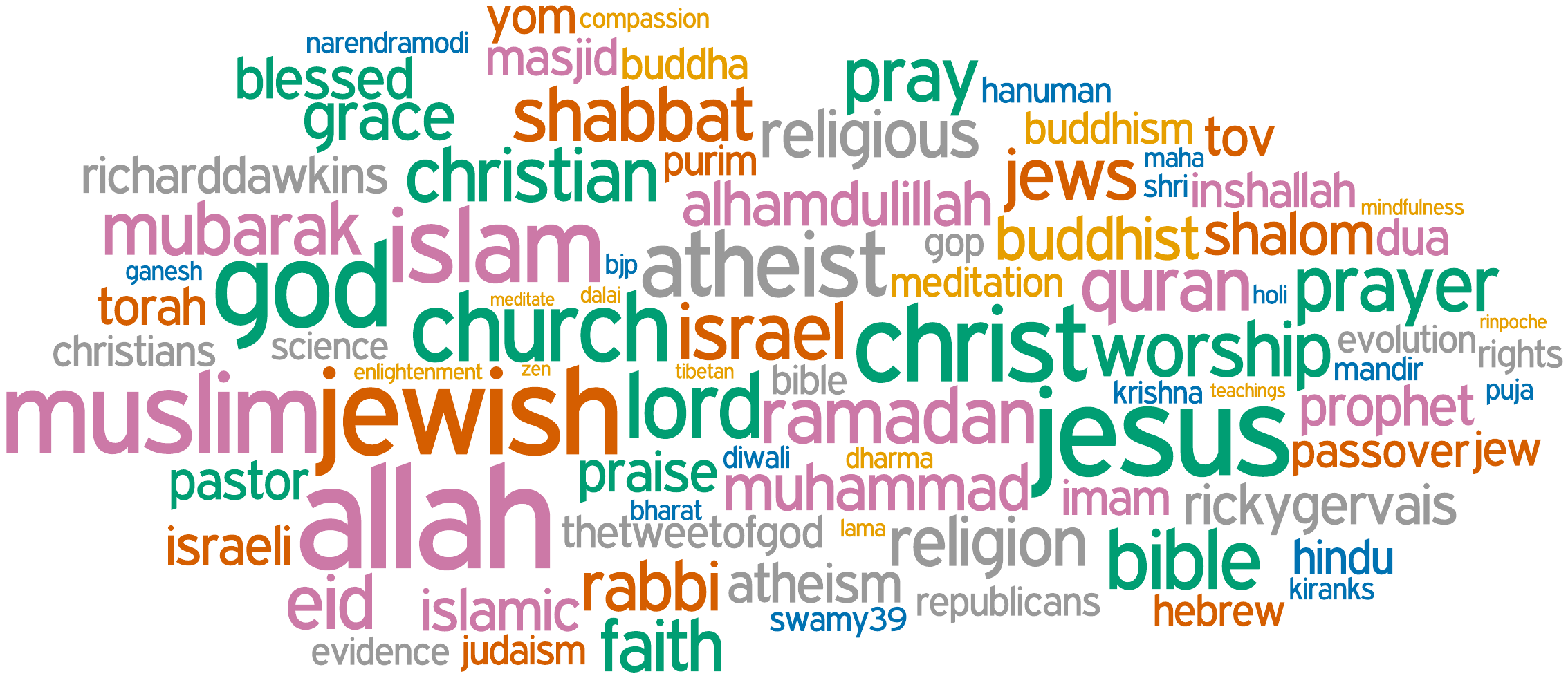}
\caption{The top 15 most discriminative words of each denomination based on a chi-square test.}
\label{wordCloud}
\vspace{-1.0em}
\end{figure*}

We first study the discriminative words in tweets that differentiate the users of one particular religious group from others by chi-square test. Specifically, we get the words from the tweet collection, and keep only the ones that appear in no less than 100 tweets. Each user group is represented by a vector of words extracted from its tweet collection, in which the words are weighted by the frequency of how many users of that group used them in their tweets (including retweets and mentions). Then a chi-square test is applied to the vector of each religious group (i.e., \emph{Atheist}, \emph{Buddhist}, \emph{Christian}, \emph{Hindu}, \emph{Jewish}, \emph{Muslim}) against the vector of the \emph{Undeclared} user group. The top 15 words that are most positively associated with each group are displayed in Figure~\ref{wordCloud}. The font size of a word in the figure is determined by its chi-square score.   

\begin{figure*}[t!]
\centering
\includegraphics[width=0.8\textwidth]{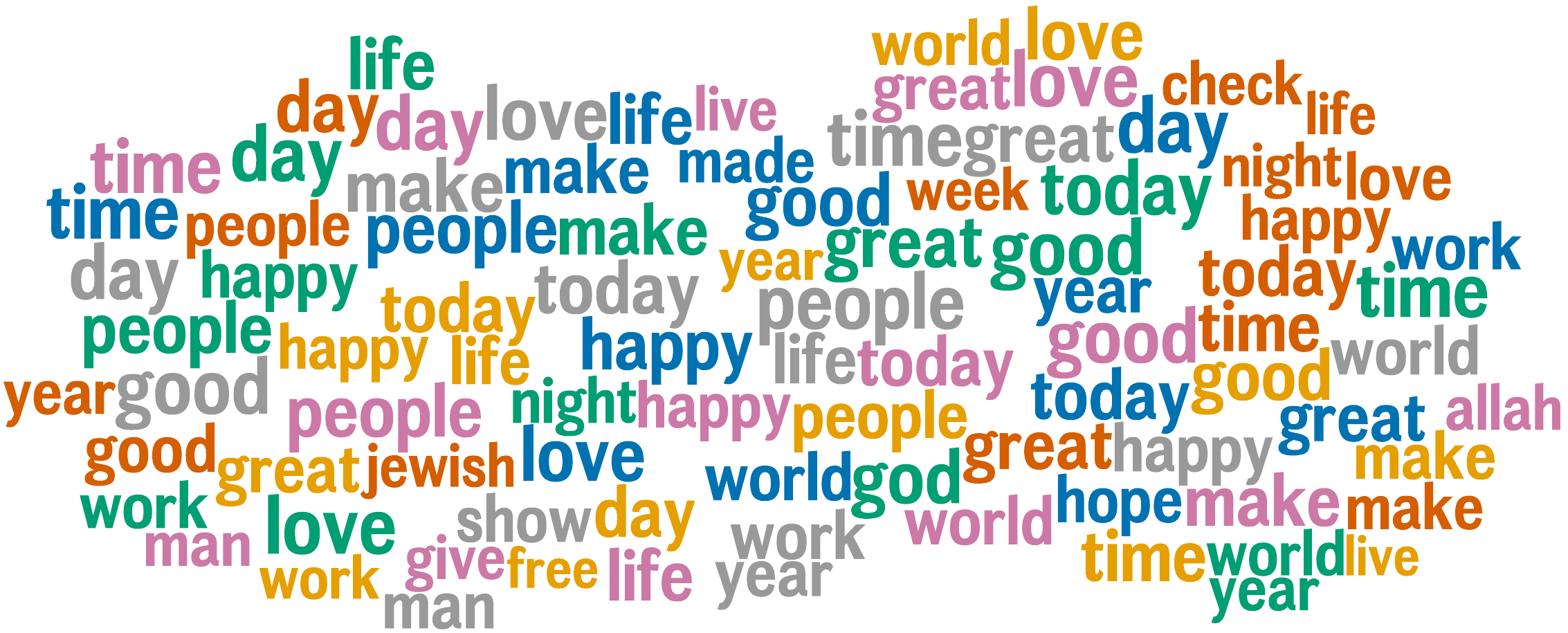}
\caption{The top 15 most frequent words for each denomination.}
\label{wordCloudFreq}
\vspace{-1.0em}
\end{figure*} 

These discriminative words are largely religion-specific, which may refer to religious images, beliefs, experiences, practices and societies of that religion. For example, the top 20 discriminative words of Christianity cover images (e.g., \textit{jesus}, \textit{god}, \textit{christ}, \textit{lord}), beliefs (e.g., \textit{bible}, \textit{gospel}, \textit{psalm}, \textit{faith}, \textit{sin}, \textit{spirit}, etc.), practices (e.g., \textit{pray}, \textit{worship}, \textit{praise}), and societies (e.g., \textit{church}, \textit{pastor}). On the other hand, Atheists show apparent preferences for topics about science (e.g., \textit{science}, \textit{evolution}, \textit{evidence}), religion (e.g, \textit{religion}, \textit{christians}, \textit{bible}) and politics (e.g., \textit{republicans}, \textit{gop}, \textit{rights}, \textit{abortion}, \textit{equality}). 

Generally, the most interesting observations relate to non-religious terms appearing as discriminative features. This includes ``evidence'' for Atheist\footnote{This is in line with recent work examining the relationship between religion and happiness on Twitter which also found Atheists to be more ``analytical'' \cite{ritteretal14spps}. Atheists are overrepresented among scientists, including top scientists (members of the Academy of Sciences) \cite{dawkins02}.}, or ``bjp'', referring to Bharatiya Janata Party\footnote{It is one of the two major parties in India, which won the Indian general election in 2014.}, for Hindu. In a sense, if our observations were to hold in a broader context, it could be seen as good for society that followers of religious groups differ most in references to religious practice and concepts, rather than in every day aspects such as music, food or other interests. This leaves more opportunities for shared experiences and culture. 
  
Whereas Figure~\ref{wordCloud} shows discriminative terms, those terms are not necessarily the most frequently used ones. Figure~\ref{wordCloudFreq} shows tag clouds that display terms according to their actual within-group frequencies. As one can see, there are lots of commonalities and terms such as ``love'', ``life'', ``people'' and ``happy'' that are commonly used by believers of all religions. This illustrates that the differences in content are not as big as Figure~\ref{wordCloud} might seem to imply. 

\subsubsection{Whom do they follow?}
\begin{figure*}[t!]
\centering
\includegraphics[width=0.8\textwidth]{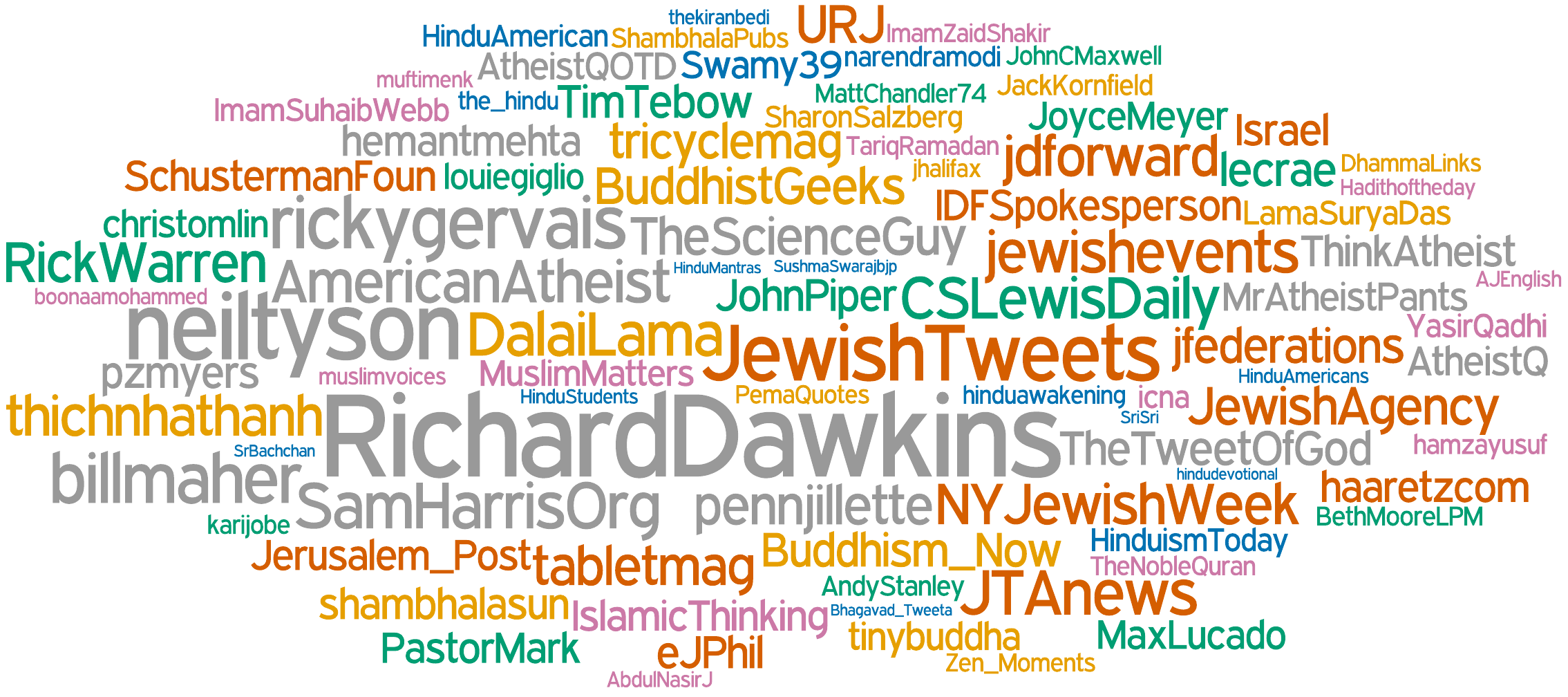}
\caption{The top 15 most discriminative Twitter accounts being followed by each denomination based on a chi-square test.}
\label{friendCloud}
\vspace{-1.0em}
\end{figure*}

We apply essentially the same methodology to study how religious people are distinguished by whom they follow on Twitter. We represent each user group by a vector of their friends, where each entry (of the vector) represents a friend being followed by the users in that group. Similar to weighting ngrams by how many users use them in the previous section, the friends in the vector are weighted by how many users from that group follow them. We then apply chi-square test to the vector of each religious group against the vector of the \emph{Undeclared} user group. Figure~\ref{friendCloud} displays the top 15 Twitter accounts (i.e., friends' screen names) that are most positively associated with each group. The font size of an account in the figure is determined by its chi-square score.

\begin{figure*}[t!]
\centering
\includegraphics[width=0.8\textwidth]{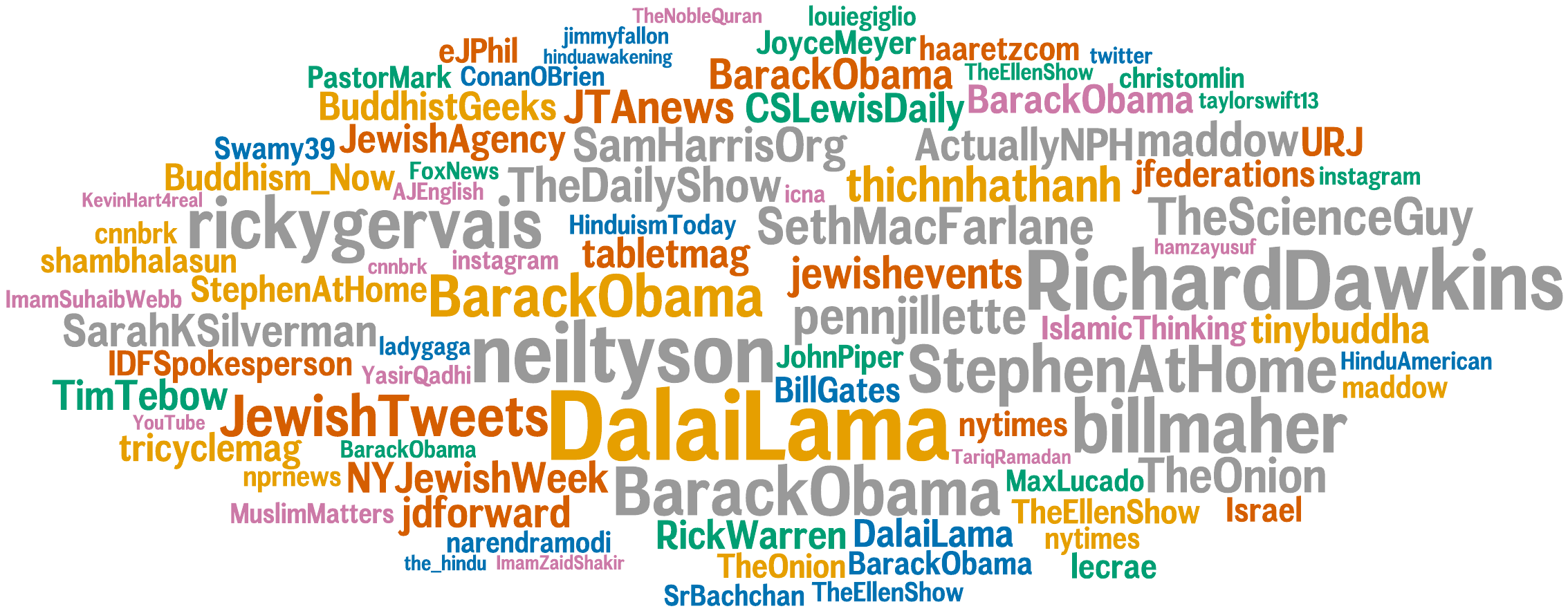}
\caption{The top 15 Twitter accounts being followed by most users of each denomination.}
\label{friendCloudFreq}
\vspace{-1.0em}
\end{figure*} 

As before, we found that the most discriminative Twitter accounts of a particular denomination are specific to that religion. E.g., \textit{IslamicThinking}, \textit{MuslimMatters}, \textit{YasirQadhi}\footnote{The Twitter account of Yasir Qadhi, who is an Islamic theologian and scholar.}, \textit{ImamSuhaibWebb}\footnote{The Twitter account of Suhaib Webb, who is the imam of the Islamic Society of Boston Cultural Center.}, and \textit{icna}\footnote{The Twitter account of Islamic Circle of North America.} are the top 5 Twitter accounts followed by Muslims which are assigned the highest chi-square scores. The top 5 Twitter accounts that characterize Atheists all belong to atheistical or irreligious celebrities, including \textit{RichardDawkins}, \textit{neiltyson}, \textit{rickygervais}, \textit{billmaher} and \textit{SamHarrisOrg}. This may have broader societal implications because these linkage or group preferences are likely to be present in the real world as well -- for instance, real world traits and behaviors such as tolerance, prejudice, and openness to experience are likely to be correlated with our findings \cite{hall10}. 

An analysis of the frequently followed users (see Figure~\ref{friendCloudFreq}) continues to show differences though and only few accounts are followed frequently by different religions. In a sense, people differ more in whom they follow rather than what they tweet about. Exceptions exist though and, for example, @BarackObama would be frequently followed by followers of most of the religions we considered.

\subsection{Religion Classification}

\begin{table*}[t!]
\centering
\scalebox{0.9}{
\begin{tabular}{llllllll}
\hline
\multicolumn{1}{|l|}{}          & \multicolumn{1}{l|}{Atheist} & \multicolumn{1}{l|}{Buddhist} & \multicolumn{1}{l|}{Christian} & \multicolumn{1}{l|}{Hindu}  & \multicolumn{1}{l|}{Jew} & \multicolumn{1}{l|}{Muslim} & \multicolumn{1}{l|}{Macro-average} \\ \hline
Tweet-based                     &                              &                               &                                &                             &                             &                             &                                    \\ \hline
\multicolumn{1}{|l|}{Precision} & \multicolumn{1}{l|}{0.747}   & \multicolumn{1}{l|}{0.6657}   & \multicolumn{1}{l|}{0.7193}    & \multicolumn{1}{l|}{0.6653} & \multicolumn{1}{l|}{0.6977} & \multicolumn{1}{l|}{0.7248} & \multicolumn{1}{l|}{0.7033}        \\ \hline
\multicolumn{1}{|l|}{Recall}    & \multicolumn{1}{l|}{0.7869}  & \multicolumn{1}{l|}{0.7388}   & \multicolumn{1}{l|}{0.7285}    & \multicolumn{1}{l|}{0.6529} & \multicolumn{1}{l|}{0.7526} & \multicolumn{1}{l|}{0.6529} & \multicolumn{1}{l|}{0.7188}        \\ \hline
\multicolumn{1}{|l|}{F1}        & \multicolumn{1}{l|}{0.7658}  & \multicolumn{1}{l|}{0.6993}   & \multicolumn{1}{l|}{0.7231}    & \multicolumn{1}{l|}{0.6588} & \multicolumn{1}{l|}{0.7241} & \multicolumn{1}{l|}{0.6868} & \multicolumn{1}{l|}{0.7097}        \\ \hline
Friend-based                    &                              &                               &                                &                             &                             &                             &                                    \\ \hline
\multicolumn{1}{|l|}{Precision} & \multicolumn{1}{l|}{0.7726}  & \multicolumn{1}{l|}{0.733}    & \multicolumn{1}{l|}{0.7681}    & \multicolumn{1}{l|}{0.7201} & \multicolumn{1}{l|}{0.7676} & \multicolumn{1}{l|}{0.7992} & \multicolumn{1}{l|}{0.7601}        \\ \hline
\multicolumn{1}{|l|}{Recall}    & \multicolumn{1}{l|}{0.8557}  & \multicolumn{1}{l|}{0.8488}   & \multicolumn{1}{l|}{0.7285}    & \multicolumn{1}{l|}{0.7148} & \multicolumn{1}{l|}{0.7595} & \multicolumn{1}{l|}{0.8351} & \multicolumn{1}{l|}{0.7904}        \\ \hline
\multicolumn{1}{|l|}{F1}        & \multicolumn{1}{l|}{0.8117}  & \multicolumn{1}{l|}{0.7864}   & \multicolumn{1}{l|}{0.7477}    & \multicolumn{1}{l|}{0.7169} & \multicolumn{1}{l|}{0.7635} & \multicolumn{1}{l|}{0.8167} & \multicolumn{1}{l|}{0.7738}        \\ \hline
\end{tabular}
}
\caption{{\footnotesize The performance of tweet-based and friend-based religiosity classification of Twitter users.}}
\label{classificationF1}
\end{table*} 

We then build classifiers to identify religious users of each denomination based on their tweet content and friend network. Specifically, we first extract a set of unigrams and bigrams (denoted as $S$) which appear in no less than 100 tweets in our tweet collection. We represent each user as a vector of unigrams and bigrams (in $S$) extracted from their tweets, where each entry of the vector refers to the frequency of that ngram in the user's tweets. The users are labeled by their denominations. We build a gold standard dataset for training and evaluating the binary classification of each denomination against the \emph{Undeclared} user group. The different sizes of the datasets affect the classification performance, e.g., the classification of Christian benefits from larger dataset. To be able to compare the performance for different denominations, we downsample the datasets of all the denominations to the same size of the Hindu dataset, the smallest one. We balance each dataset to contain the same number of positive and negative instances. For each religious group, we train the SVM classifiers using LIBLINEAR \cite{fan2008liblinear}, and apply 10-fold cross validation to its dataset. Similarly, we also represent each user as a vector of their friends, where each entry of the vector refers to whether the user follows a user X ($1$ - if the user follows X, and $0$ - otherwise.) For each denomination, we build the gold standard dataset, balance it, train the SVM classifiers, and estimate the performances by 10-fold cross validation.

Table~\ref{classificationF1} reports the results. The tweet-based classification achieves a macro-average F1 score of $0.7097$, and the friend-based classification achieves a macro-average F1 score of $0.7738$. It demonstrates the effectiveness of content features and network features in classifying Twitter users' religiosity, and network features appear to be superior to content features. According to the F1 score, the difficulty level of recognizing a user from a specific religious group based on their \textit{tweet content} is (from easiest to hardest): Atheist $<$ Jew $<$ Christian $<$ Buddhist $<$ Muslim $<$ Hindu, while the difficulty level of recognizing a user from a specific religious group based on their \textit{friend network} is (from easiest to hardest): Muslim $<$ Atheist $<$ Buddhist $<$ Jew $<$ Christian $<$ Hindu.

\section{Linkage Preference}

In this section, we focus on exploring ingroup and outgroup relations. We construct four directed networks based on religious users following (friend), being-followed-by (follower), mentioning, and retweeting others, respectively. Following and being-followed-by relations are extracted from users' friend and follower lists, respectively. Mention and retweet relations are extracted from tweets, i.e., whether user A retweeted at least one tweet from user B, and whether user A mentioned user B in at least one of his/her tweets, respectively. Here we do not separate reply from mention. If a tweet addresses a specific user by including ``@'' followed by the user's screen name and it is not a retweet (e.g., marked with ``RT''), we call it a mention.

For each user in our dataset, we count the numbers of all his/her connections (i.e., friends, followers, retweets, or mentions) and the connections with each religious group. Then we calculate the proportions of his/her ingroup (same-religion) connections and the connections to users from other groups. We get the average proportions of ingroup and outgroup connections for each group by adding that proportions of all the users in the group together and dividing by the number of users. The raw proportion may not reflect the linkage preference since it is affected by the number of users in a group. The connections to Christians may always account for the biggest proportion because there are much more Christians than others in the dataset and even random linkage would give the illusion of preferring connection to Christians. So in addition to the raw proportion, we also estimate the expected proportion of connections to a specific user group by the fraction of users of a certain religion in a random user sample. 

To be specific, in Section~\ref{dataCollection} we describe how we generate random numbers as Twitter user IDs, and collect user profiles from Twitter by these IDs. From all the valid U.S.\ user profiles collected in this way, we identify the users included in any religious denominations from our sample, and get the proportion of users of each denomination as the expectation of how likely a Twitter user connects with a user from a certain group. The expected proportions of connections are 0.0466\% (Atheist), 0.0259\% (Buddhist), 1.3358\% (Christian), 0.0013\% (Hindu), 0.0207\% (Jew) and 0.0414\% (Muslim). Note that these proportions are low as the vast majority of Twitter users do not explicitly state a religious affiliation in their profile. We then use the relative difference of the proportion to its expected value to represent the linkage preference. For example, Christian-Christian following accounts for $4.33\%$ of all followings of a Christian user in average, and its relative difference compared to the expected value is $\frac{4.33\%-1.3358\%}{1.3358\%} = 2.2$. These values are often referred to as ``lift'' in statistics.

We observe a preference for religious users to connect to others that share the similar belief to them, e.g., religious users are much more likely to follow other users of the same religion than of a different religion. For example, the same-religion followings of Hindu account for a proportion of $0.99\%$ and the relative difference is $737.3$, and same-religion followings of Jews account for a proportion of $8.15\%$ and the relative difference is $392.3$. Overall, following someone of the same religion is $646$ times as likely as following someone of a different religion based on a macro-average of six denominations, if we estimate the following likelihood with the relative proportion obtained by dividing the raw proportion by the expected value.

\begin{figure}[ht!]
\centering
\includegraphics[width=0.48\textwidth]{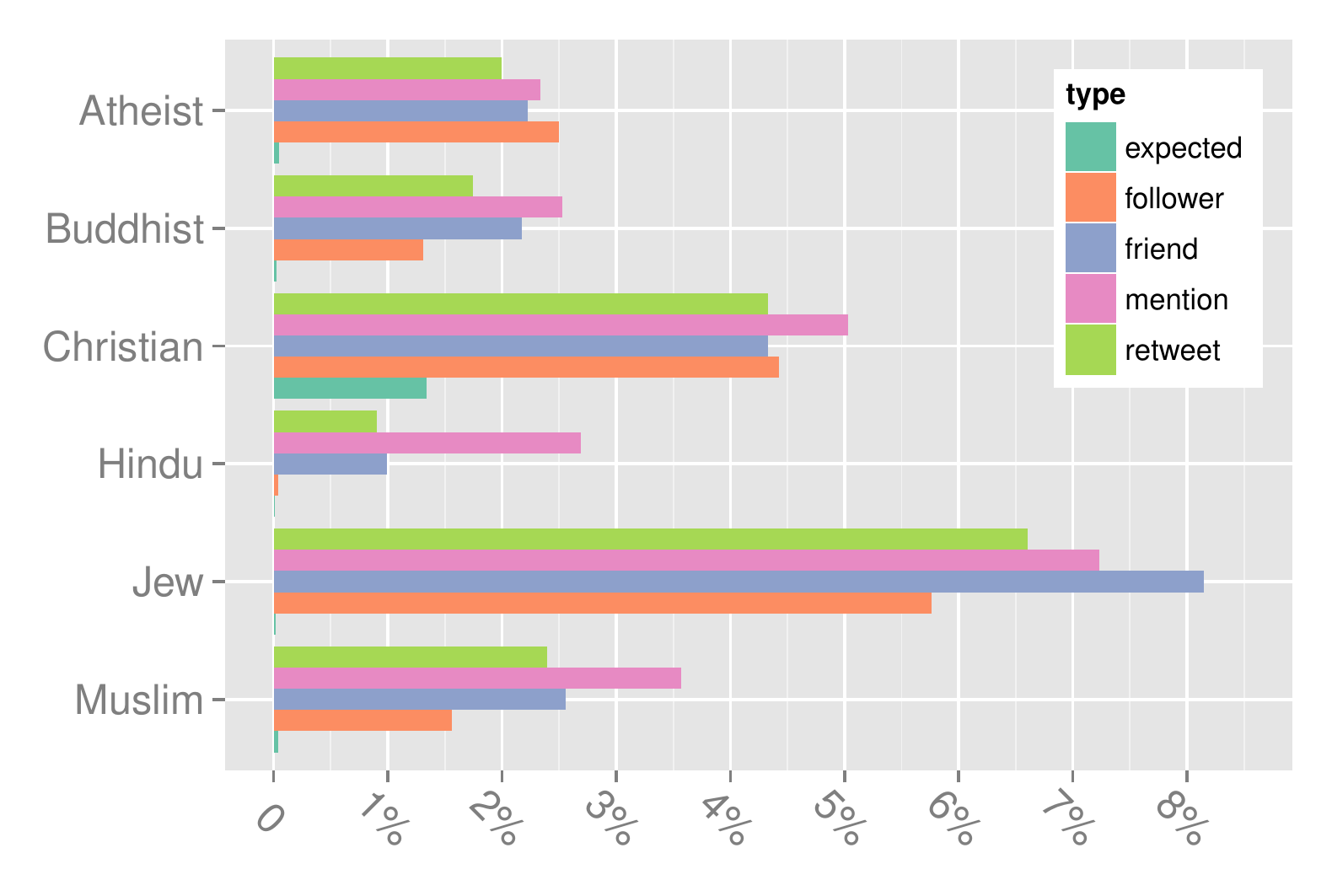}
\caption{The proportion of same-religion relations of each religious group.}
\label{sameReligionRelations}
\vspace{-1.0em}
\end{figure} 

Figure~\ref{sameReligionRelations} plots the proportions of same-religion relations of different types. We compute the average proportions (per user) of being-followed-by, retweet and mention in the same way as we compute that of following. The expected proportions of connections are the same as we have described in the previous section. The same-religion linkage preference exists in all types of connections across all the religious groups. However, because our analysis is conducted on the users who self-reported their religious affiliations, it is probably biased toward very religious users, and for the other religious users who do not disclose their religion/belief on Twitter, the ingroup linkage preference may not be as strong.

\begin{figure}[ht!]
\centering
\includegraphics[width=0.48\textwidth]{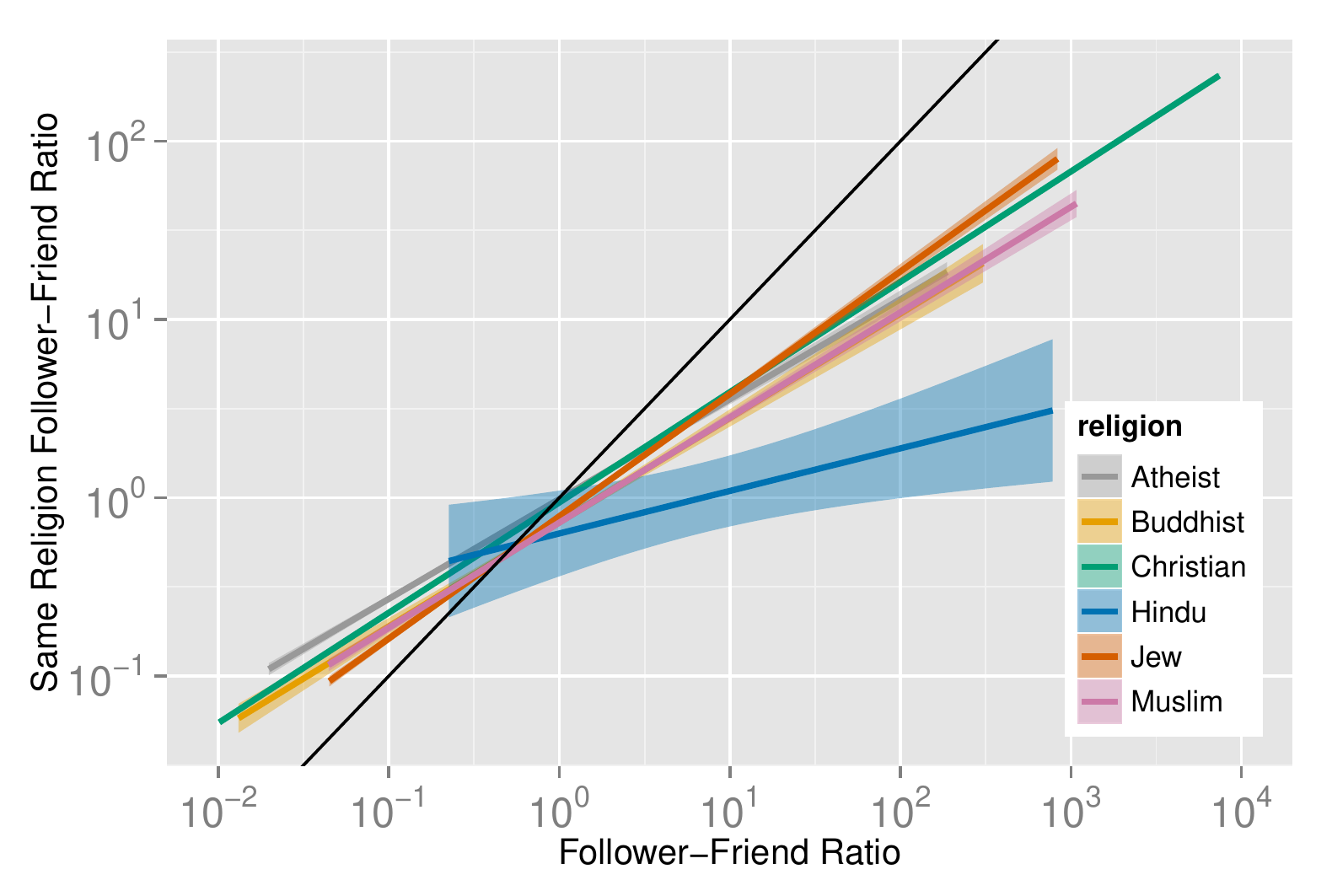}
\caption{Follower-friend ratio vs.\ same-religion follower-friend ratio. Linear smoothing is applied.}
\label{tff}
\vspace{-1.0em}
\end{figure}

From Figure~\ref{sameReligionRelations} we observe such preference is stronger in the friend network than in the follower network for many religious groups such as Muslim, Jew, and Hindu. Note that this is at first sight paradoxical as when A follows B of the same religion this means that B is followed A by the same religion.\footnote{Some readers might rightly think of the somewhat related ``Friendship Paradox'' that your friends or followers have more friends and followers than you \cite{hodas2013friendship}.} In order to explain this phenomenon, we plot the follower-friend ratio against the same-religion follower-friend ratio of the users in each group in Figure~\ref{tff}. It shows that the same-religion linkage preference of follower network is diluted by the out-group followers of the users who have more followers than friends. The ratio of same-religion followers of a local priest (e.g., placing at the bottom-left area in the coordination) may be higher than that of the same-religion friends, while the pope (e.g., placing at the top-right area in the coordination) may have many out-group followers that dilutes the ratio of same-religion followers. When the users in the top-right area contribute more to the overall proportions, the average ratio of same-religion friends is higher than that of the same-religion followers, otherwise, it is lower. 

\section{Conclusion}

In this paper we used data from more than 250k U.S.\ users to describe the religious landscape on Twitter. We showed the distribution of Twitter users with a self-declared religious affiliation is a reasonable match to the distribution of religious believers according to surveys. We then characterized how different religions differ in terms of the content of the tweets and who they follow, and show that for the task of telling random users from religious users, a user's friends list provides more effective features than the content of their tweets. We find and quantify proof of within-group linkage preference for following, being-followed, mentioning and retweeting across all of our religions. 

The ultimate goal of studies such as ours is not to study religion \emph{on Twitter}, but to study religion \emph{per se}, and arguably it will be more and more feasible in the future. Because more and more communication happens online, also more religious communication is likely to happen online. A key feature of religion is replication, and communication is key for such replication. Religion is a replicator -- it replicates itself, its dogma, longitudinally (from generation to generation) and horizontally (across population), and in that sense it relies heavily on transmission media, Twitter being one of them.
 
Twitter might be replaced by ``the next big thing'' but religion itself will not disappear in the foreseeable future, though it is continuously evolving along with the cultural context it is embedded in. To ensure a broader relevance of studies using online data it is therefore important to validate findings through separate channels and to ground research in existing literature. Online data can serve well to form hypotheses related to group formation, emotional stability or demographic correlates such as race or income and to guide follow-up studies looking at more holistic data and root causes. 

There are several limitations and at the same directions for future research. For example, in our current analysis, we only used the content of tweets to discover and describe discriminative tokens. No efforts were made to detect differences in dimensions such as sentiment or mood or other linguistic dimensions. In future work, we hope to gain clues as to what makes a religion stand out, e.g., when it comes to providing emotional stability or dealing with personal setbacks.

%\section{Acknowledgments}

% Balancing columns in a ref list is a bit of a pain because you
% either use a hack like flushend or balance, or manually insert
% a column break.  http://www.tex.ac.uk/cgi-bin/texfaq2html?label=balance
% multicols doesn't work because we're already in two-column mode,
% and flushend isn't awesome, so I choose balance.  See this
% for more info: http://cs.brown.edu/system/software/latex/doc/balance.pdf
%
% Note that in a perfect world balance wants to be in the first
% column of the last page.
%
% If balance doesn't work for you, you can remove that and
% hard-code a column break into the bbl file right before you
% submit:
%
% http://stackoverflow.com/questions/2149854/how-to-manually-equalize-columns-
% in-an-ieee-paper-if-using-bibtex
%
% Or, just remove \balance and give up on balancing the last page.
%
%\balance

% If you want to use smaller typesetting for the reference list,
% uncomment the following line:
% \small
\bibliographystyle{acm-sigchi}
\bibliography{chen_US-Religious-Landscape-on-Twitter}
\end{document}